\documentclass[epj]{svjour}

\usepackage{graphics, epsfig, euscript}

\def\A{{\EuScript A}}
\def\Q{{\EuScript Q}}
\def\X{{\EuScript X}}

\begin{document}

\title{\boldmath Non-Valence Fock States in Heavy-to-Light Form Factors 
                 at Large Recoil}

\author{Bj\"orn O.~Lange \hfill {\tt CLNS 03/1845}}

\institute{
Newman Laboratory for Elementary-Particle Physics, 
Cornell University, Ithaca, NY 14853, U.S.A.}

\date{Received: date / Revised version: date}

\abstract{
We studied three-particle Fock state contributions to heavy-to-light form 
factors in the context of soft-collinear effective theory and found that they 
enter at leading power. These contributions are non-factorizable due to the 
appearance of endpoint singularities, however they do not violate spin-symmetry
relations at leading power. In this talk I present their numerical estimation 
in a crude model in which the ``soft overlap'' contribution is cut off and 
find that they might lower the standard values for the form factors at maximum 
recoil significantly. Furthermore I briefly discuss the role of soft-collinear 
messenger modes in the region of soft overlap.
\PACS{{12.39.St}{Factorization} \and
      {13.20.He}{Decays of bottom mesons}} 
} 

\authorrunning{Bj\"orn O.~Lange}
\titlerunning{Non-Valence Fock States in Heavy-to-Light Form Factors 
              at Large Recoil}

\maketitle

\section{Introduction}
The calculation of many $B$-decay amplitudes can be significantly simplified 
using QCD factorization theorems \cite{Beneke:1999br}, and soft-collinear 
effective theory (SCET) \cite{SCET} provides an excellent environment to 
discuss such factorization proofs. It is of great interest to apply this 
technology to $B\to$ {\em light meson} form factors, since they enter the 
calculation of amplitudes for many important rare decays such as 
$B\to K^* \gamma$, $B\to \pi \pi$, etc. At large recoil the leading 
contributions to the 10 standard form factors can be described by only three 
independent universal functions $\zeta_p, \zeta_\parallel, \zeta_\perp$, due 
to the spin-symmetry of highly energetic fermions \cite{Charles:1998dr}. These 
relationships are, however, not exact, and an interesting question is whether 
corrections to them can be written in a factorized form, as proposed by Beneke 
and Feldmann \cite{Beneke:2000wa,Beneke:2003xr}
\begin{equation}
f_i (q^2) = C_i \,\zeta(E) \; + \; \phi_B \otimes T_i \otimes \phi_L 
            \;+\; \ldots
\label{eq1}
\end{equation}

Here, $\phi_B$ and $\phi_L$ denote leading order light-cone distribution 
amplitudes (LCDAs) of the $B$ meson and the light recoiling meson, 
respectively. The coefficients $C_i$ and the kernel functions $T_i$ are 
perturbatively calculable and the ellipses denote power suppressed 
contributions. The symbols $\otimes$ denote a convolution integral, which 
emerges through spectator interactions. Note that this formula does not 
define the functions $\zeta(E)$ unambiguously and that factorizable 
contributions to the form factor $f_i$ can be associated with either term 
in (\ref{eq1}) as long as they obey the spin-symmetry relations. 

Part of any proof of factorization must address the question of convergence of 
the convolution integrals, as well as the question whether two-particle Fock 
states dominate the amplitude at leading power. The main motivation for our 
study was to find examples in which this is not the case or for which answers 
to these questions are non-trivial.

In this talk I discuss the role of three-particle Fock state contributions to 
the form factors \cite{prep}. They are found to appear in leading power and are
non-factorizable due to endpoint singularities. However, they do not violate 
spin-symmetry and can therefore be associated with the first term in equation
(\ref{eq1}).

\section{Calculation}

\begin{figure}[b!]
\begin{center}
\epsfig{file=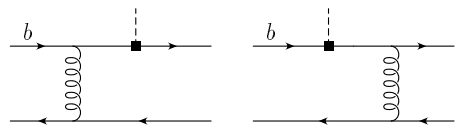, width=6cm}
\caption{Spectator interaction via exchange of a hard-collinear gluon.}
\label{fig:twopart}
\end{center}
\end{figure}

The goal of this calculation is to match the QCD amplitude 
$\langle L|\bar \psi_q \Gamma \psi_b| B \rangle$, where $L$ represents a light
pseudoscalar or vector meson and $\Gamma$ an appropriate Dirac structure,
onto the corresponding SCET operator matrix elements. These operators contain 
at least one heavy quark and one light soft quark in the initial state and 
two collinear quarks in the final state. We calculated three different 
classes of diagrams that include spectator interactions: the two-particle to 
two-particle amplitude, as shown in Fig. \ref{fig:twopart}, the two-particle 
to three-particle amplitude, in which the operators contain one extra final 
collinear gluon, and the three-particle to two-particle amplitude with one 
extra soft gluon in the initial state. We obtain contributions to the matching 
calculation by attaching an extra gluon anywhere in either of the two QCD 
diagrams in Fig. \ref{fig:twopart} as long as this leads to an additional 
off-shell propagator. These off-shell modes are either hard or hard-collinear 
and will be integrated out in the matching procedure. Diagrams that involve 
emission of an extra collinear gluon from the heavy quark line lead to power 
suppression; however, the remaining diagrams yield leading power contributions 
to the form factors, which then involve three-particle Fock state LCDAs of the 
final light meson after projecting onto the physical states. The same can be 
observed for diagrams involving three partons of the $B$ meson. The results 
for each of the three classes include terms that diverge logarithmically at 
the endpoints, such as, for example, 
\begin{equation}
\int\limits_0^\infty d\omega \; \frac{\phi_+^B(\omega)}{\omega^2} 
\hspace{4mm} {\rm or} \hspace{4mm}
\int\limits_0^1 dx \; \frac{\phi_\pi(x)}{(1-x)^2} \; .
\label{convint}
\end{equation}
It is convenient to introduce light-cone coordinates, in which case $\omega$ 
can be identified with the ``plus-component'' of the soft spectator momentum, 
and $x$ denotes the fraction of the ``minus-component'' of the pion momentum 
carried by the collinear quark emerging from the weak interaction. The 
integrals in (\ref{convint}) diverge as $\omega$ or $\bar x=1-x$ become 
smaller than the originally assigned scaling in powers of $\lambda=\Lambda/m_b$
in SCET. For comments see section \ref{sec:4}. 

Let me stress here that these endpoint singularities arise in a direct 
matching of $QCD \to SCET$. In a two-step matching procedure $QCD \to SCET_I
\to SCET_{II}$, where the intermediate theory lives on a hybrid scale 
$\sqrt{\Lambda m_b}$ and the final theory coincides with $SCET$, it has been 
argued \cite{Bauer:2002aj} that no divergences appear in the first matching 
step. We must therefore conclude that the second matching step is non-trivial 
in that factorizable terms in $SCET_I$ do not necessarily match onto 
factorizable terms in $SCET_{II}$.

\section{A rough numerical estimation}
Another interesting aspect of our analysis arises from the fact that 
three-particle contributions to the final light meson appear at leading 
power. In this section we would like to estimate their relative numerical 
impact on the three universal form factors, which we define, in accordance with
\cite{Beneke:2000wa}, to coincide with the standard form factors
\begin{equation}
\zeta_P = f_+ \quad , \quad \zeta_\parallel = A_0 
\quad , \quad \zeta_\perp=T_1.
\end{equation}
We regularize divergent integrals, such as (\ref{convint}), by introducing 
cutoffs $\omega_0 = \epsilon_h \lambda_B$, $x_0 = \epsilon_l$ which separate 
the contributions near (``soft overlap'') and away from the endpoints. The
HQET parameter $1/\lambda_B$ is defined as the first inverse moment of 
$\phi_+^B$ and provides the necessary mass dimension to $\omega_0$. Our 
belief is that the situation near the endpoints can not be correctly 
described by soft and collinear degrees of freedom only, but rather 
necessitates the introduction of a ``soft-collinear'' mode 
\cite{Becher:2003qh,Becher:2003kh}. 
The discussion of their role is left for section \ref{sec:4}. Here 
we simply set these contributions to zero and study the dependence of the 
regulated convolution integrals to the cutoffs $\epsilon_l$ and $\epsilon_h$, 
keeping in mind that these cutoffs serve as a transition mark between soft 
($\omega \sim \lambda$) and soft-collinear modes ($\omega \sim \lambda^2$), 
or collinear ($\bar x \sim 1$) and soft-collinear ($\bar x \sim \lambda$) 
modes. Typical values of these cutoffs are around $(\epsilon_l, \epsilon_h) 
\approx (0.1,0.3)$. 

\begin{figure}[t!]
\begin{center}
\epsfig{file=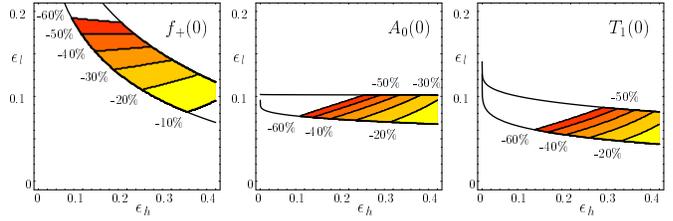, width=8.8cm}
\caption{Visualization of the relative size of three-particle Fock state 
contributions to the form factors $f_+(0)$, $A_0(0)$ and $T_1(0)$. 
The bands
reflect the regions in parameter space in which all terms that cannot be 
associated with a three-particle configuration of the light final meson give 
rise to the values found by LCSR within a 30\% error. These bands are 
color-coded by the value of the ratio of three-particle and two-particle 
contributions.}
\label{fig:numerics}
\end{center}
\end{figure}

Since we would like to compare relative contributions from three-particle 
configurations of the final light meson we define two quantities for each 
form factor: $\triangle F_i{(2p)}$ denotes all terms involving purely 
two-particle final states, i.e.\ in the pseudoscalar case $\phi_\pi$ and the 
two-particle part of the twist-three wave-function $\phi_p$. Correspondingly 
$\triangle F_i{(3p)}$ collects terms of three-particle origin. In Fig. 
\ref{fig:numerics} we show regions in the $\epsilon_l - \epsilon_h$ plane 
in which the $\triangle F_i{(2p)}$ correspond to values predicted by 
light-cone sum rules (LCSR) \cite{Ball:1998kk,Ball:1998tj}. This is
reasonable since the LCSR analysis finds a negligible numerical impact from 
three-particle configurations, quoting \cite{Ball:1998kk}: {\em ``putting all 
intrinsic higher-twist parameters [$\ldots$] to zero, the form factors change 
by at most 3\%''}. The band width denotes a generous 30\% error on the LCSR 
central values. We color-coded these bands with the ratio 
$\triangle F_i{(3p)}/\triangle F_i{(2p)}$. 

The immediate observation is that the three-particle contributions are negative
for all reasonable values of $(\epsilon_l, \epsilon_h)$ and hence lower the 
numerical values for the form factors. Furthermore their impact is significant 
in the vicinity of $(\epsilon_l, \epsilon_h) \approx (0.1,0.3)$ and might be, 
with the exception of $f_+$, as large as 50\%. 

Although the above model serves only as a very crude estimate, one must wonder 
how this result compares with the method of LCSR. The main difference between 
the Factorization/SCET approach and the LCSR approach lies in the sensitivity 
to the light-cone structure of the $B$ meson through heavy-to-light currents 
that are smeared over the light-cone in the former case and the interpolation 
of the $B$ meson by local currents in the latter. Furthermore the continuum 
model adopted in LCSR effectively cuts off any contributions near the endpoints
$\bar x \to 0$ in a rough way. The question whether the method of LCSR 
reliably captures the effects of spectator interactions has recently been 
addressed in \cite{Ball:2003bf}. In this work it was pointed out that one can 
probe the light-cone structure of the $B$ meson by invoking a second sum rule 
for the (process independent) HQET parameter $1/\lambda_B$, albeit missing 
finer (process dependent) details related to spectator interactions.

\section{Endpoint configurations}
\label{sec:4}
The existence of endpoint singularities suggests that the phase-space below the
cutoffs $\epsilon_l$ or $\epsilon_h$, the soft-overlap region, obtains no 
suppression. If we describe the soft spectator momentum in light-cone 
coordinates it scales like $(\lambda, \lambda, \lambda)$, whereas the 
collinear momentum scales like $(\lambda^2, 1, \lambda)$. However, considering 
the situation in which $\omega$ becomes smaller, i.e.\ below the cutoff, the 
soft mode naively changes into $(\lambda, \lambda, \lambda) \to 
(\lambda^2, \lambda, \lambda)$. Similarly, in the endpoint $\bar x \to 0$ the 
collinear scaling changes into $(\lambda^2, 1, \lambda) \to 
(\lambda^2, \lambda, \lambda)$. In SCET, these configurations are described by
on-shell fields whose momenta scale like $(\lambda^2, \lambda, \lambda^{3/2})$.
This mode, called ``soft-collinear mode'', has recently been studied in 
\cite{Becher:2003qh,Becher:2003kh} using the method of regions. 
In the case of form factors we can construct box and pentagon diagrams and 
show that soft-collinear exchanges lead to leading loop-momentum regions.

Note that in this region of phase-space the formally hard-collinear gluon 
interacting with the spectator now becomes soft or collinear, and is thus no 
longer integrated out when matched onto SCET operators. Since we can choose 
the external states to be eigenstates of the leading order SCET Lagrangian we 
may allow for one more unsuppressed soft or collinear gluon exchange 
between the upper and lower fermion line in Fig. \ref{fig:twopart} to form a 
box or pentagon diagram with only soft or collinear external legs. 

However, there is no need to keep these soft-collinear modes in the theory, 
because we do not allow for such external momenta and thus require no source 
terms in the path integral. It is therefore possible to integrate them out, 
leaving an induced non-local interaction between soft and collinear fields 
which symbolically takes the form
\begin{equation}
S_{SC} = \int d^4x \, d^4y \, \frac{\triangle(x_- y_+)}{(x_- y_+)^2} \; 
\bar \X \A_c[x_-] \A_s \Q[y_+] \; .
\label{eq:4}
\end{equation}

The explicit form can be found in \cite{Becher:2003qh}. In this notation $\X$ 
and $\A_c$ denote gauge invariant collinear quark and gluon fields, $\Q$ 
and $\A_s$ denote gauge invariant soft quark and gluon fields 
\cite{Hill:2002vw}, and $\triangle$ is a soft-collinear propagation function.
Power counting shows that this interaction scales like $\lambda^{3/2}$ and 
therefore enters the form factor calculation at leading power \cite{prep}.

Let me summarize: at leading power in SCET we find four types of operators 
which appear in the tree-level matching of QCD heavy-to-light form factors. The
first three types are ``local'' operators (except for a light-cone separation) 
containing two+two, three+two and two+three soft+collinear fields. In the soft
overlap region we have in addition to them a time-ordered product of a 
heavy-to-collinear current with the induced soft-collinear interaction 
(\ref{eq:4}).

\section{Conclusion}
Heavy-to-light form factors at large momentum transfer are terrific (and 
terrifying) examples for processes in which factorization is problematic. Two- 
and three-particle Fock states contribute at leading power and display endpoint
singularities. We studied the first effect numerically by disregarding the soft
overlap and introducing cutoff regulators that render the convolution integrals
finite. It was found that three-particle contributions might lower the 
numerical values of the form factors significantly, especially in the 
$B \to V$ case, which is also favored by the analysis of radiative decays 
$B \to V \gamma$ in the framework of factorization 
\cite{Bosch:2001gv,Beneke:2001at}. Our calculation also supports the finding 
by Luo and Rosner \cite{Luo:2003hn} who extrapolated the $q^2$ dependence of 
pseudoscalar form factors from lattice data into the low $q^2$ region and 
favor rather low values . Lastly I discussed soft-collinear messenger modes 
and endpoint singularities.

\subsection*{Acknowledgments}
This work has been performed in collaboration with Farrukh Chishtie and 
Matthias Neubert, to whom I am greatly indebted. I also would like to thank 
M. Beneke, T. Feldmann, S. Bosch, E. Lunghi and S. Descotes-Genon for many 
interesting discussions during this conference.

\end{document}